\documentclass[twocolumn,amsmath,amssymb,superscriptaddress,nofootinbib,longbibliography,pra]{revtex4-2}

\usepackage{graphicx}
\usepackage{dcolumn}
\usepackage{braket}

\renewcommand\vec{\mathbf}

\usepackage[version=3]{mhchem} 

\begin{document}

\title{Circumventing the polariton bottleneck via dark excitons in 2D semiconductors} 

\author{Jamie M. Fitzgerald}
\affiliation{Department of Physics, Philipps-Universität Marburg, 35032, Marburg, Germany}

\author{Roberto Rosati}
\affiliation{Department of Physics, Philipps-Universität Marburg, 35032, Marburg, Germany}

\author{Beatriz Ferreira}
\affiliation{Department of Physics, Chalmers University of Technology, 412 96, Gothenburg, Sweden}

\author{Hangyong Shan}
\affiliation{Institute of Physics, Carl von Ossietzky University, Oldenburg, 26129, Germany}

\author{Christian Schneider}
\affiliation{Institute of Physics, Carl von Ossietzky University, Oldenburg, 26129, Germany}

\author{Ermin Malic}
\affiliation{Department of Physics, Philipps-Universität Marburg, 35032, Marburg, Germany}

\date{\today}

\begin{abstract}
Efficient scattering into the exciton polariton ground state is a key prerequisite for generating Bose-Einstein condensates and low-threshold polariton lasing. However, this can be challenging to achieve at low densities due to the polariton bottleneck effect that impedes phonon-driven scattering into low-momentum polariton states. The rich exciton landscape of transition metal dichalcogenides (TMDs) provides potential intervalley scattering pathways via dark excitons to rapidly populate these polaritons. Here, we present a microscopic study exploring the time- and momentum-resolved relaxation of exciton polaritons supported by a \ce{MoSe2} monolayer integrated within a Fabry-Perot cavity. By exploiting phonon-assisted transitions between momentum-dark excitons and the lower polariton branch, we demonstrate that it is possible to circumvent the bottleneck region and efficiently populate the polariton ground state. Furthermore, this intervalley pathway is predicted to give rise to, yet unobserved, angle-resolved phonon sidebands in low-temperature photoluminescence spectra that are associated with momentum-dark excitons. This represents a distinctive experimental signature for efficient phonon-mediated polariton-dark-exciton interactions. 
\end{abstract}

\maketitle

Transition metal dichalcogenide (TMD) monolayers support tightly bound excitons with binding energies in the hundreds of meV \cite{wang2018colloquium,perea2022exciton}. Alongside optically accessible bright excitons \cite{chernikov2014exciton}, these 2D semiconductors also exhibit a rich variety of dark excitonic states. In particular, intervalley excitons, where the constituent electron and hole are located in different valleys of the hexagonal Brillouin zone, possess a large centre-of-mass momentum which forbids direct radiative excitation and recombination \cite{malic2018dark}. These momentum-dark states can be efficiently populated via exciton-phonon scattering on a femtosecond timescale \cite{selig2016excitonic,brem2019intrinsic,schmitt2022formation}, meaning that the energetic ordering of the exciton landscape plays a crucial role in determining the relaxation dynamics and photoluminescence (PL) \cite{selig2018dark, brem2020phonon}.

The combination of a large exciton binding energy and oscillator strength makes TMDs excellent candidates for room-temperature polaritonics \cite{dufferwiel2015exciton,zhang2018photonic,zhao2023exciton}. Exciton polaritons are hybrid light-matter states that arise when an excitonic material is integrated within an optical cavity, and the light-exciton coupling strength exceeds all material and photonic dissipation channels \cite{savona1999optical, kavokin2003thin}. Recently, exciton–polariton condensation at cryogenic temperatures \cite{anton2021bosonic} and room-temperature polariton lasing \cite{zhao2021ultralow} have been demonstrated for TMDs. These important milestones highlight the growing need to understand how exciton relaxation is modified for 2D semiconductors within the strong coupling regime over a wide range of temperatures. Alongside polariton-polariton \cite{zhao2022nonlinear} and polariton-electron \cite{hartwell2010numerical} interactions, polariton-phonon scattering is one of the main mechanisms driving polariton relaxation, and even dominates at low densities. Previously, we have found that the substantial photonic character and greatly reduced density of states of the lower polariton (LP) branch within the lightcone drastically suppresses acoustic-phonon-driven intravalley scattering in TMDs by two orders of magnitude \cite{ferreira2022microscopic}, in line with earlier studies on conventional semiconductors \cite{tassone1997bottleneck,tartakovskii2000relaxation}. Combined with the short radiative lifetime of polaritons, this leads to a severe reduction in emission at small angles and is known as the "polariton bottleneck effect" \cite{tassone1997bottleneck,tartakovskii2000relaxation}. The bottleneck has significant implications for the performance of polariton-based lasing devices \cite{imamog1996nonequilibrium,liew2011polaritonic}, as well as fundamental studies on polariton condensates \cite{deng2010exciton}. In both cases, a mechanism is needed to populate the ground state at a faster rate than the radiative lifetime to achieve the required minimum polariton occupation. Vibrationally assisted polariton relaxation has been demonstrated to be an effective strategy to suppress the bottleneck effect at elevated temperatures. For organic semiconductors this tends to be particularly efficient due to a broad energetic range of available vibrational modes \cite{coles2011vibrationally}. While for inorganic \cite{boeuf2000evidence,maragkou2010longitudinal} and crystalline hybrid semiconductors \cite{laitz2023uncovering}, the LP branch can be tuned such that the energy separation between its minimum and the bottleneck reservoir coincides with an optical phonon energy. This strategy has been demonstrated to reduce the lasing threshold for a quantum well polariton laser \cite{maragkou2010longitudinal}. Further approaches for boosting polariton relaxation involve using spatial confinement \cite{paraiso2009enhancement}, and polariton-polariton scattering at higher densities \cite{tartakovskii2000relaxation,zhao2022nonlinear}. 

\begin{figure}[t!]
\includegraphics[width=\columnwidth]{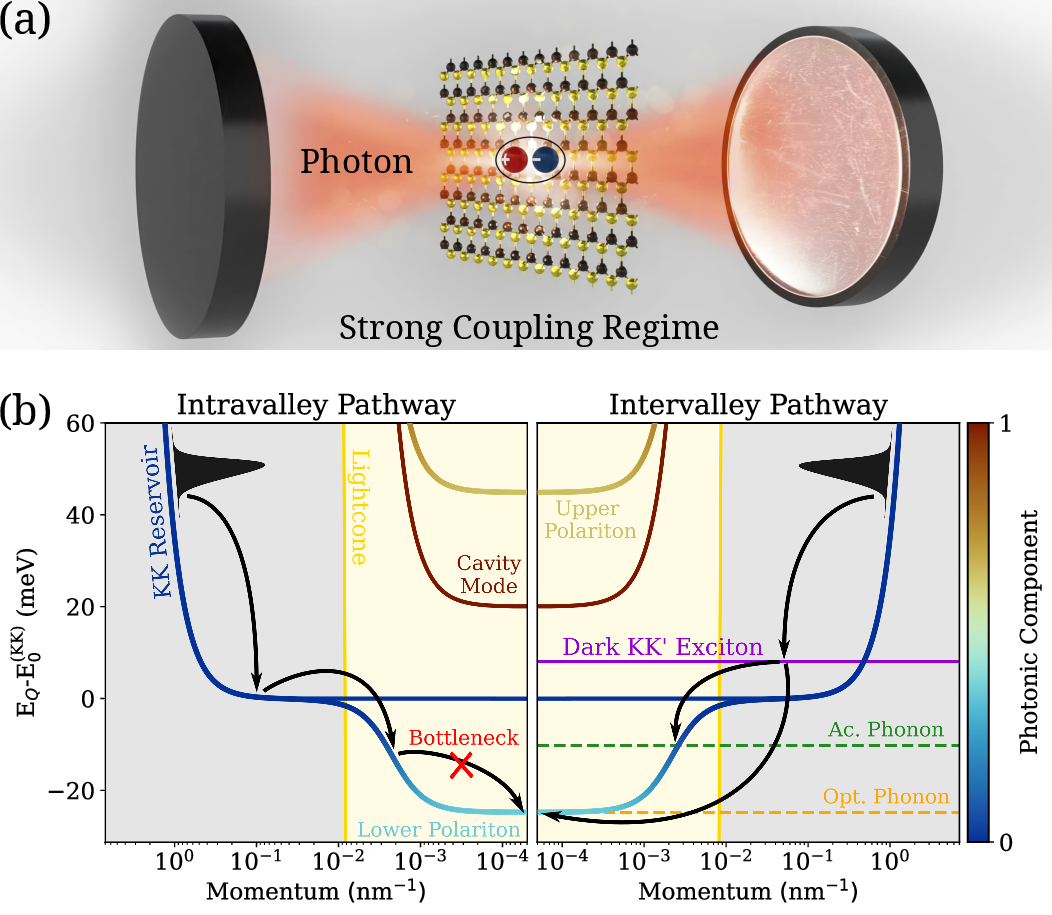}
\caption{(a)~Illustration of a TMD monolayer integrated within a Fabry-Perot microcavity in the strong coupling regime. (b)~Schematic of possible relaxation processes (black arrows) for the lower branch of a molybdenum-based exciton polariton. The left-hand side shows the bottleneck effect for intravalley scattering via acoustic phonons. The right-hand side shows the possible intervalley relaxation processes via acoustic (dashed green line) and optical (dashed orange line) phonon scattering from dark KK' excitons (solid purple line). For this particular cavity detuning ($\Delta=20$ meV), the $\vec{Q}=0$ polariton state is exactly one optical phonon energy lower than the bottom of the KK' valley.} \label{fig:fig_1}
\end{figure}

The implications of the full exciton landscape in TMDs has yet to be understood in the context of exciton polariton relaxation. Recently, the important role of dark exciton states in TMD polaritonics was demonstrated by using the Rabi splitting to push the LP below a dark excitonic state in a \ce{WSe2} monolayer, leading to a brightening of PL via the increased occupation of the LP branch \cite{shan2022brightening}. Using a Wannier-Hopfield approach \cite{Fitzgerald2022,konig2023interlayer}, we explore in this work how exciton polaritons supported by a \ce{MoSe2} monolayer integrated within a Fabry-Perot microcavity relax towards the lightcone (Fig. \ref{fig:fig_1}(a)). In particular, we highlight how phonon-assisted scattering from the momentum-dark exciton can bypass the bottleneck region and allows for the efficient population of low-momentum states in the LP branch at room temperature (Fig.~\ref{fig:fig_1}(b)). This leads to highly detuning-dependent and material-specific relaxation dynamics based on the energy separation of the LP branch and the dark excitons. Furthermore, we show that at low temperatures a unique, to-date unobserved, phonon sideband signature appears in PL. Distinct from the bare exciton case \cite{brem2020phonon,rosati2020temporal}, these sidebands are angle-resolved and specific to molybdenum-based TMDs within the strong coupling regime.

\begin{figure*}[t!]
\includegraphics[width=\textwidth]{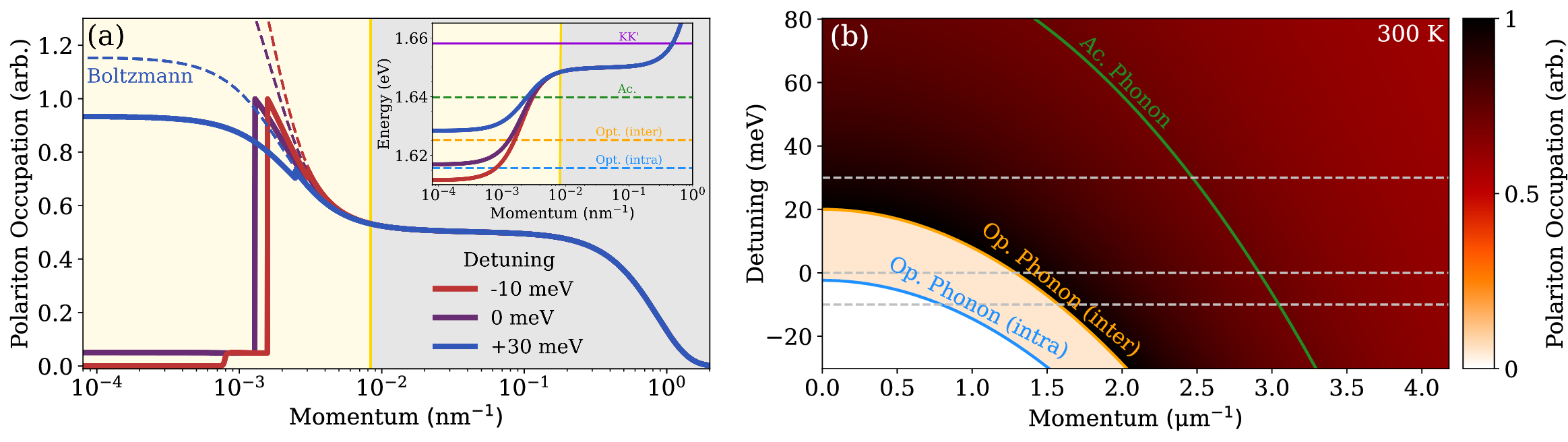}
\caption{(a)~Lower polariton steady-state occupation for three representative detunings at a temperature of $300$ K (solid lines), and the corresponding thermalized Boltzmann distributions (dashed lines). The inset shows the polariton dispersion for each detuning. (b)~Detuning dependence of the steady-state polariton occupation within the lightcone. The orange and green lines show the opening of optical and acoustic phonon scattering channels from the KK' exciton valley, respectively. The blue line shows the opening of the intravalley optical phonon channel.
\label{fig:fig_2}}
\end{figure*}

\section{Results}
\textbf{Bypassing the polariton bottleneck:}
Detuning a microcavity via the cavity length (i.e., mirror separation, see Fig.~\ref{fig:fig_1}(a)) modifies the dispersion and light-exciton composition of the LP branch within the lightcone (Fig.~\ref{fig:fig_1}(b)). This impacts both the strength of polariton-phonon scattering \cite{lengers2021phonon,ferreira2022microscopic}, and the relative position of bright and dark states within the energy landscape \cite{shan2022brightening} (the latter are unmodified by the presence of the microcavity). As a result, significant control over the temporal evolution of the polariton population is gained. To explore this, we apply a material-specific Wannier-Hopfield method \cite{Fitzgerald2022,konig2023interlayer} to model exciton polaritons supported by an hBN-encapsulated \ce{MoSe2} monolayer integrated in the center of a symmetric $\lambda/2$ Fabry-Perot cavity (Fig.~\ref{fig:fig_1}(a)). We consider only the lowest 1s excitons associated with the KK, KK' and K$\Lambda$ states, where the first and second letter denote the reciprocal-space valley in which the respective Coulomb-bound hole and electron are localized \cite{malic2018dark}, respectively. Solving the Wannier equation reveals that \ce{MoSe2} is a direct semiconductor, where the dark KK' excitons (solid purple line in Fig.~\ref{fig:fig_1}(b)) lie about $10$ meV above the bottom of the bright KK exciton ($E_{0}^{\text{KK}}$, solid blue curve) \cite{selig2018dark}. In the context of bare exciton optics and dynamics, KK' excitons in \ce{MoSe2} monolayers play only a minor role: they lead to a small linewidth increase in linear optical spectra at elevated temperatures due to phonon absorption \cite{selig2016excitonic}. In stark contrast, within the strong coupling regime, KK' excitons are surprisingly essential in determining the polariton occupation within the lightcone as they can provide an additional exciton reservoir via phonon-induced intervalley population transfer.

Due to their short radiative lifetime, a correct description of polariton thermalization requires time-resolved simulations. The dynamics of the incoherent polariton occupation of the $n$th branch, at the center-of-mass momentum $\vec{Q}$, is found to be governed by the momentum-dependent semiclassical Boltzmann equation \cite{savona1999optical}
\begin{equation}
\dot{N}_{n\vec{Q}}(t) =
\sum_{m,\vec{Q}'}W_{m\vec{Q}',n\vec{Q}} N_{m\vec{Q}'}(t)
-2\tilde{\Gamma}^{\text{out}}_{n\vec{Q}} N_{n\vec{Q}}(t), \label{eq:boltzmann}
\end{equation}
where $\tilde{\Gamma}^{\text{out}}_{n\vec{Q}}$ is the total out-scattering rate, which is the sum of the polariton radiative decay, $\gamma_{n\vec{Q}}$, and the phonon dephasing rate $\Gamma_{n\vec{Q}}$. Furthermore, $W_{m\vec{Q}',n\vec{Q}}$ is the corresponding in-scattering matrix describing the rate of scattering from the state $\ket{m,\vec{Q}'}$ to $\ket{n,\vec{Q}}$ via both phonon absorption and emission. Crucially, we include the full momentum and valley dependence of the in- and out-scattering rates, allowing us to explore the role of the exciton landscape on polariton relaxation. We initialise all simulations with a Gaussian distribution centered at $50$ meV in the LP branch to mimic non-resonant excitation \cite{savona1999optical,kavokin2003thin}. Further details on the theoretical approach can be found in the Methods and Supplementary Material (SM). 

The steady-state momentum-resolved polariton occupation, $N_{\vec{Q}}(t_\infty)$, is shown in Fig.~\ref{fig:fig_2}(a) for three representative detunings at room temperature. Outside of the lightcone (denoted by the vertical gold line), the polaritons closely follow a thermalized Boltzmann distribution, $N^0_{\vec{Q}}$ (dashed lines), for all three detunings. We will refer to these purely excitonic states as the "KK exciton reservoir" (see Fig.~\ref{fig:fig_1}(b)). In contrast, the steady-state occupation within the lightcone is strongly detuning-dependent. It can be expressed as
\begin{equation}
N_{n\vec{Q}}(t_\infty)= \frac{\Gamma_{n\vec{Q}}}{\Gamma_{n\vec{Q}}+\gamma_{n\vec{Q}}}N^{0}_{n\vec{Q}}, \label{eq:steady_state}
\end{equation}
which can be derived from the static limit of Eq.~(\ref{eq:boltzmann}) assuming that the in-scattering is dominated from states outside the lightcone (see SM). The prefactor quantifies the ratio of polaritons that are scattered back into the exciton reservoir to be recycled, relative to the total out-scattering rate. The latter includes polaritons that escape from the cavity via mirror leakage. While both the phonon-driven out-scattering and radiative decay rate are sensitive to the tunable exciton-light composition, the former also crucially depend on the available scattering channels that are allowed by energy-momentum conservation.

For a red-detuned cavity ($\Delta = E^{\text{cav}}_0-E^{\text{exc}}_0=-10$ meV, red line), the LP at low momenta within the lightcone is more photon-like (i.e., a sharper dispersion, see red curve in the inset of Fig. \ref{fig:fig_2}a) and possesses a negligible occupation. This exemplifies the bottleneck effect and highlights the reluctance of polaritons to thermalize in the absence of strong scattering channels, i.e., there is a large deviation between the calculated steady-state occupation and a Boltzmann distribution (red solid vs dashed line in Fig. \ref{fig:fig_2}a) due to the suppression of intravalley acoustic phonon scattering within the lightcone \cite{ferreira2022microscopic}. For increasing momenta within the lightcone, the occupation shows three sharp step-like increases before tending towards the Boltzmann distribution. This is indicative of the opening of specific phonon scattering channels \cite{ferreira2022signatures}. The first small jump in occupation at around $8\times10^{-4}$ nm$^{-1}$ is the opening of the intravalley optical phonon scattering channel. Above this particular momentum, the energy-momentum conservation can be fulfilled and optical phonons (with an energy of $34$ meV \cite{jin2014intrinsic}) can scatter from the KK exciton reservoir into the lightcone \cite{boeuf2000evidence,maragkou2010longitudinal}. The other two sharp jumps are a consequence of the opening of intervalley scattering channels from the KK' exciton reservoir via optical ($33$ meV) and acoustic ($18$ meV) phonons \cite{jin2014intrinsic} at $Q=1.6\times10^{-3}$ and $3\times10^{-3}$ nm$^{-1}$, respectively. In particular, the larger jump in occupation stems from the optical phonon-driven intervalley scattering. 

At zero detuning ($\Delta = 0$, purple line in Fig. \ref{fig:fig_2}a), there is a small increase in occupation at $\vec{Q}=0$ as the intravalley optical phonon channel is now open over all possible LP momenta within the lightcone (i.e., the purple dispersion is above the dashed blue line in the inset), but there is still a substantially depleted population compared to the thermalized limit (purple dashed line). The two step-like increases from intervalley scattering are still present, but slightly shifted towards lower momenta due to the shallower polariton dispersion (purple curve, inset). The occupation within the lightcone changes drastically in a blue-detuned cavity ($\Delta =+30$ meV, blue line). There is now only one small step-like increase due to the opening of the acoustic phonon channel, and the occupation follows a slightly depleted Boltzmann distribution within the lightcone. This is because optical phonon-driven scattering from KK' excitons is now possible at all momenta (see blue curve, inset), revealing that momentum-dark excitons can act as an additional reservoir to efficiently populate the entire LP branch at elevated temperatures.

These observations cannot be explained by the detuning-induced change in the light-exciton nature of the polariton (i.e., Hopfield coefficients), as the changes in the population are very sharp in momentum. To further illustrate this point, in Fig.~\ref{fig:fig_2}(b) we show the steady-state LP occupation against momentum and detuning within the lightcone. There are three distinct regions, which are separated by the opening of intra- (blue line) and intervalley (orange line) optical phonon scattering channels, as well as the opening of the intervalley acoustic channel (green line). The momentum at which the energy-momentum conservation can be satisfied for an intervalley scattering process towards the LP depends on the detuning, exciton reservoir energy, and the associated phonon energy. In the case of the KK' exciton reservoir, this can be stated as $E_Q^{\text{P}}(\Delta) = E_Q^{\text{KK'}} - E^{\text{ph}}$, and is indicated by the green and orange lines in Fig.~\ref{fig:fig_2}(b) for acoustic and optical phonons, respectively. Efficient population of the $\vec{Q}=0$ state via the KK' exciton reservoir is activated for a blue-detuned cavity of $20$ meV. This occurs when the lowest LP state coincides in energy with the bottom of the KK' valley minus the optical phonon energy (see schematic Fig.~\ref{fig:fig_1}(b)). In the SM we present a detuning study where all intervalley channels are artificially switched off. There we find a severely reduced occupation within the lightcone at all detuning values underlining the importance of these channels for bypassing the polariton bottleneck.

\begin{figure}[t!]
\includegraphics[width=\columnwidth]{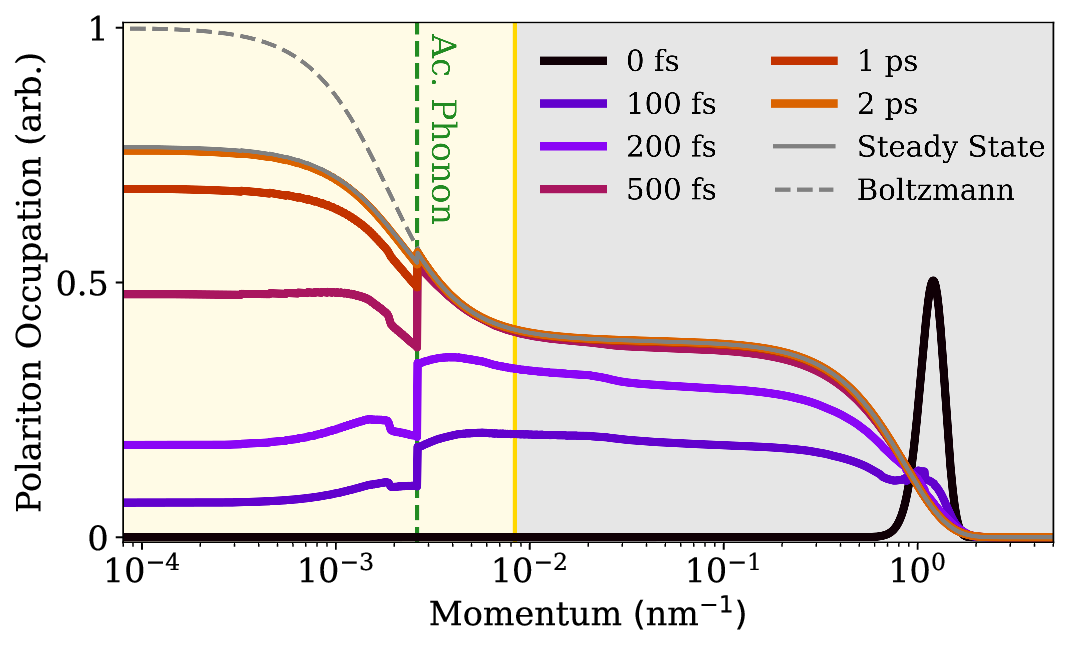}
\caption{Snapshots at fixed times of the lower polariton occupation at $300$ K and a cavity detuning of $\Delta=20$ meV. This particular detuning corresponds to the opening of the optical intervalley scattering channel for the polariton ground state.
\label{fig:fig_3}}
\end{figure}

\textbf{Time-resolved occupation:} 
The solution of the polaritonic Boltzmann equation provides microscopic access to the temporal evolution of polariton states. Figure.~\ref{fig:fig_3} illustrates the dynamics of  the momentum-resolved LP occupation for a cavity detuned to $\Delta=20$ meV, which corresponds to the opening of the optical phonon scattering channel (cf. Figs.~\ref{fig:fig_1}(b)). The KK reservoir outside the lightcone thermalizes within approximately $500$ fs (solid and dashed grey lines coincide) via the efficient phonon-assisted scattering at room temperature, in agreement with previous studies of bare excitons in \ce{MoSe2} \cite{selig2018dark}. This highlights that the dynamics outside of the lightcone can be understood with a purely excitonic picture. States within the lightcone that can be populated via both acoustic and optical phonon-driven scattering from the KK' reservoir, thermalize on a similar timescale as the KK exciton reservoir. In contrast, the low-momenta polaritons within the lightcone reach a steady-state (solid grey line given by the solution of Eq.~\ref{eq:steady_state}) on a longer timescale of about two ps. Interestingly, there is a significant rise in the polariton population already within $100$ fs, which could have relevance for ultrafast polaritonic devices \cite{liew2011polaritonic}. The crossover between these two regions of the LP dispersion is indicated by the large kink (vertical dashed green line) in the occupation. The size of this jump decreases over time as the occupation tends towards the quasi-thermalized limit of a depleted Boltzmann distribution. 

The rapid build-up of occupation in the LP ground state is a direct consequence of the quick transfer of population from the initially populated KK exciton reservoir to KK' excitons over tens of fs via phonon emission (see SM for the dynamics of KK' excitons). Given the elevated temperature and close energetic separation ($<10$ meV) between the KK and KK' exciton reservoirs, the latter hold nearly half the total population at steady-state. Significantly, the large occupation of the KK' exciton at high temperatures is not of a transient nature and does not depend on specific excitation conditions. This means it can act as an efficient exciton reservoir for the low-momenta LP states over tens or hundreds of ps. Overall, our results highlight that the dynamics of the LP branch can be divided into two domains. States outside the lightcone thermalize rapidly and independently of whether the system is in the strong coupling regime. In contrast, the states within the lightcone are predominately populated via a phonon-assisted in-scattering from the thermalized exciton reservoirs. Depending on the cavity detuning, this can efficiently bypass the bottleneck region and populate low-momenta polariton states.

\begin{figure}[t!]
\includegraphics[width=\columnwidth]{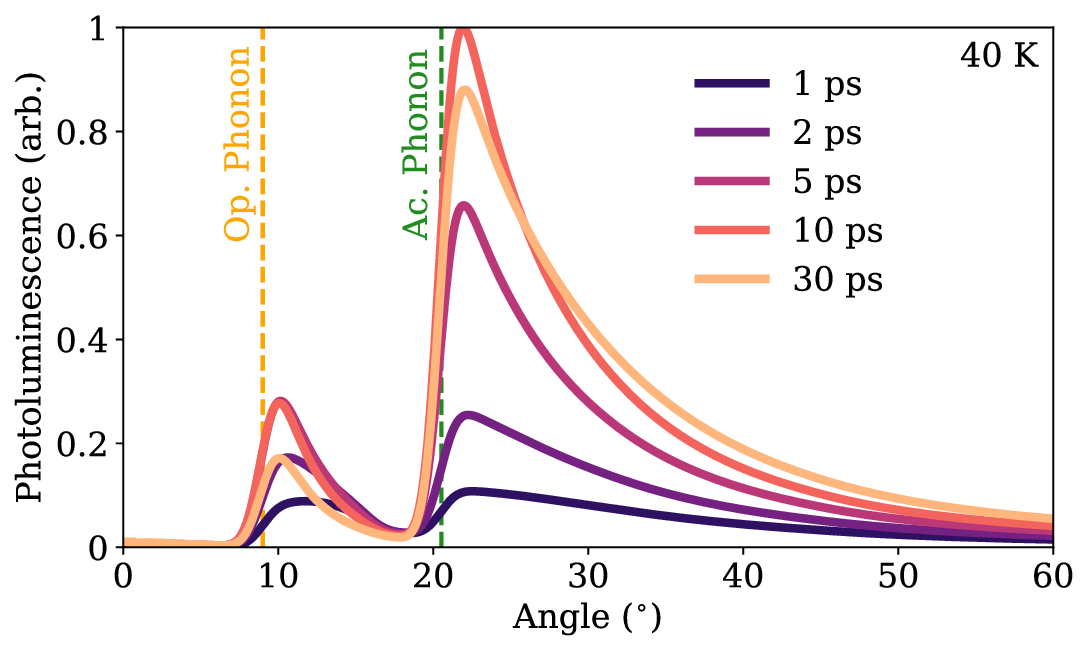}
\caption{Snapshots of the angle-resolved photoluminescence (PL) at a temperature of $40$ K for a zero-detuned cavity. The PL is evaluated at the polariton energy along the lower branch dispersion. An integration over a $2^\circ$ detection window is performed to mimic realistic experimental measurements.
\label{fig:fig_4}}
\end{figure}

\textbf{Phonon-assisted polariton photoluminescence:} At low temperatures, phonon-assisted relaxation at all detunings is inefficient as the polariton radiative decay is larger than the total in-scattering rate (see Eq.~\ref{eq:steady_state}). Because polaritons do not thermalize within the lightcone, scattering pathways from the KK' reservoir imprint a distinctive signature on angle- and time-resolved \emph{resonant} PL, i.e., evaluated at the polariton energy. This is shown in Fig.~\ref{fig:fig_4} for a zero-detuned system at $40$K. The resonant polariton PL can be expressed as (see SM):
\begin{equation}
    I_{\vec{Q}}^{\mathrm{res}}(t) = \frac{\gamma_{n\vec{Q}}}{\gamma_{n\vec{Q}}+\Gamma_{n\vec{Q}}} N_{n\vec{Q}}(t), \label{eq:PL}
\end{equation}
where the prefactor quantifies the momentum-dependent ratio of the photon leakage rate out of the cavity and the total out-scattering. Note that there is a direct correspondence between the polariton in-plane momentum and the angle of PL emission, $\sin(\theta)=\hbar cQ/E_Q^{\text{P}}$. We find an enhanced PL at emission angles corresponding to the opening of phonon scattering channels (dashed vertical lines in Fig.~\ref{fig:fig_4}). Given the low temperature, the peaks at $10^\circ$ and $20^\circ$ are of a transient nature in the time range shown here. They grow in amplitude before peaking around 10-20 ps, and then slowly decay towards the steady state. This is a consequence of the KK' exciton valley being initially overpopulated via intervalley transfer from the KK exciton reservoir, before then slowly thermalizing over tens of ps. The difference in PL peak height stems from the occupation, and is a consequence of a larger acoustic exciton-phonon matrix element and the increased excitonic character of the LP at larger momenta. The asymmetric lineshape with a long decaying tail is reminiscent of phonon-assisted PL, and is related to the kinetic energy distribution of the source exciton reservoir \cite{snoke1987quantum}. 

In tungsten-based TMDs, momentum-indirect excitons are energetically lower than the bright KK exciton and therefore carry the majority of the excitonic population at low temperatures. This large occupation results in pronounced phonon sidebands red-shifted by the phonon energy from the dark exciton energy \cite{brem2020phonon,rosati2020temporal}. These sidebands corresponds to a weak momentum-conserving process where a momentum-indirect exciton can recombine and emit a virtual photon within the lightcone via the simultaneous interaction with a phonon. Here, we observe a similar process, but with the virtual photon replaced with a polariton state. This has two major impacts: (i) The mechanism is more efficient as it is now a resonant process with a real final state. (ii) While the usual phonon-assisted PL lineshape appears as a function of frequency, here it corresponds to the resonant PL evaluated along the LP dispersion (Eq.~\ref{eq:PL}). This means it must be resolved with respect to the angle of emission, which can be shifted with cavity detuning (see SM for a detuning study of the LP occupation at $40$ K). These angle-dependent PL peaks represent a new hallmark of polariton-phonon interactions. To date, only a Purcell enhancement of phonon sidebands has been predicted in cavity systems \cite{bottge2012enhancement}. Note that this phonon-assisted polariton PL is contingent on the dark exciton energy being on the order of the exciton-light coupling strength above the bright exciton energy. This is satisfied for KK' excitons in molybdenum-based TMDs. In contrast, this process will be challenging to observe with a tungsten-based TMD as tuning the polariton to typical phonon energies below the KK' or K$\Lambda$ excitons will lead to extremely light-like polaritons. This is in stark contrast to the bare excitonic case where phonon sidebands are only visible in tungsten-based TMDs \cite{brem2020phonon}. These findings are a promising way to utilize strong coupling physics as a probe of phonon-driven intervalley transitions.

In conclusion, we have demonstrated the crucial importance of the full exciton landscape, including dark and bright states, for polariton relaxation dynamics in TMD monolayers. At room temperature, we demonstrate that momentum-dark excitons can provide an efficient reservoir to populate the ground state and bypass the polariton bottleneck. At low temperatures, they lead to unique phonon sidebands visible in angle-resolved PL that can be controlled by cavity detuning. These signatures in the polariton PL provide an experimental tool to study the strength of the intervalley polariton-phonon scattering, and potentially even measure the energy of momentum-dark excitons. Overall, our work provides microscopic insights into polariton dynamics in atomically thin semiconductors, and has technological relevance for the design of compact and low-threshold polariton lasers that operate at room temperature. \\

We acknowledge funding from the Deutsche Forschungsgemeinschaft
(DFG) via SFB 1083 and the regular project 524612380.

\section{Methods}
 Exciton energies, wavefunctions, and oscillator strengths are obtained microscopically by solving the intervalley Wannier equation \cite{berghauser2014analytical,selig2016excitonic}, with DFT input used to characterise the two-band parabolic approximation for the electronic bandstructure \cite{kormanyos2015k}. The end mirror reflectance is set to $99.8\%$ (quality factor of $1570$), which corresponds to a realistic, high-quality microcavity \cite{zhao2021ultralow}. The exciton-light coupling is found to be $33$ meV for a $\lambda/2$ cavity. In order to model the strong coupling between the bright KK exciton and the cavity photon, we utilize the Hopfield diagonalisation, which provides access to polariton energies and Hopfield coefficients \cite{Fitzgerald2022}. The latter, $V_{\mu n,\vec{Q}}/U_{n\vec{Q}}$, give a measure of the $\mu$th excitonic/photonic character of polariton state $\ket{n,\vec{Q}}$. In particular, the polariton-phonon scattering strength is given by the exciton-phonon matrix element \cite{selig2016excitonic}, $D_{\mu\nu,q}$, scaled by the excitonic component of both the initial and final polariton state: \cite{lengers2021phonon,ferreira2022microscopic,ferreira2022signatures} $\tilde{D}_{nm,\vec{Q},\vec{Q}'}= \sum_{\mu,\nu} V_{\mu n,\vec{Q}}^{*}D_{\mu\nu,|\vec{Q}'-\vec{Q}|}V_{\nu m,\vec{Q}'}$. These matrix elements describe the strength of intravalley scattering within the polariton branches, as well as intervalley scattering between dark excitons and polaritons \cite{ferreira2022signatures}. To this end, it is convenient to consider the KK' and K$\Lambda$ excitons as dark polaritons with a $100\%$ excitonic component. Long-range acoustic phonons are treated within the Debye (first-order) approximation, while short-range acoustic and all optical phonons are treated in the Einstein (constant energy) approximation. Using a deformation potential approximation, the exciton-phonon matrix elements are calculated with phonon energies and electron-phonon coupling strengths obtained from DFT calculations \cite{jin2014intrinsic}, as detailed in previous studies \cite{selig2018dark,brem2020phonon}. 

The polariton Boltzmann equation (Eq. (\ref{eq:boltzmann})) is a specification of the polaritonic Bloch equations in the purely incoherent limit \cite{takemura2016coherent}. It can be derived using an equation of motion approach \cite{kira2006many}, and assuming a low polariton density and thermalized bath of incoherent phonons. It has been successfully used to study exciton polariton dynamics in a range of different material systems, such as conventional quantum wells \cite{tassone1997bottleneck,hartwell2010numerical}. The phonon-driven out-scattering rate is determined by summing over all possible energy- and momentum-conserving channels
\begin{align}
\Gamma_{n\vec{Q}}=& \frac{\pi}{\hbar}\sum_{m\Vec{Q}',\alpha,\pm}
\left|\tilde{D}_{\alpha,n\vec{Q},m\vec{Q}'} \right|^{2}
\left(\frac{1}{2}\pm\frac{1}{2}+n_{\alpha,|\vec{Q}-\vec{Q}'|}^{\text{ph}}\right) \nonumber
\\ 
&\times \delta\left(E_{m\vec{Q}'}^{\text{P}}-E_{n\vec{Q}}^{\text{P}} \pm E^{\text{ph}}_{\alpha,|\vec{Q}-\vec{Q}'|}\right). \label{eq:phonon_rate}
\end{align}
Here, $n^{\text{ph}}$ represents the phonon occupation, approximated by a Bose-Einstein distribution, and the summation is taken over phonon emission/absorption ($\pm$) and all relevant phonon modes, $\alpha$. The polariton radiative decay for a symmetric cavity of length $L$ and mirror transmission $T$ is given by $\gamma_{n\vec{Q}}=cT|U_{n\vec{Q}}|^2/(2L)$ \cite{Fitzgerald2022}.

\bibliography{bib}

\renewcommand{\thesection}{S.\Roman{section}}
\renewcommand{\theequation}{S.\arabic{equation}}
\renewcommand{\thefigure}{S\arabic{figure}} 

\onecolumngrid

\section*{Supplementary materials}

\subsection{Computational Details}

\subsubsection{Wannier Equation}

The Wannier equation is solved for the bright KK as well as the momentum-dark KK' and K$\Lambda$ 1s excitons \cite{berghauser2014analytical}. The latter have a negligible contribution to the exciton dynamics as they lie $\sim 140$ meV above the bright KK exciton \cite{selig2018dark}. Spin-dark excitons are disregarded throughout this work as they require a spin-flip process to be occupied and will only be important on longer timescales of tens to hundreds of ps \cite{rosati2020temporal}. The spectral position of the KK exciton in hBN-encapsulated \ce{MoSe2} is fixed to $1.65$ eV according to experimental PL measurements \cite{ajayi2017approaching}. The screened Coulomb potential is modelled as a generalised Keldysh potential \cite{brem2019intrinsic} using the dielectric constants for the \ce{MoSe2} monolayer obtained by density functional theory ($\epsilon_\perp=7.2$ and $\epsilon_\parallel=16.8$) \cite{laturia2018dielectric}, and $\epsilon_{\text{sub}} = 4.5$ for the hBN layers.

To estimate the optical matrix element, $M_0$, which dictates the strength of the exciton-photon coupling and consequentially the Rabi splitting, a two-band $\vec{k}\cdot\vec{p}$ expansion is used \cite{haug2009quantum},  $M_0=m_{0}\sqrt{E_g/(4m_{\text{r}})}$, where $m_{\text{r}}=m_{\text{e}}m_{\text{h}}/(m_{\text{e}}+m_{\text{h}})$ is the reduced electron-hole mass, $m_{0}$ is the free electron mass, and $m_{\text{e}/\text{h}}$ is the effective electron/hole mass. This gives an estimate of $1/(2\gamma^{\text{exc}})=0.3$ ps for the exciton radiative lifetime, where $\gamma^{\text{exc}}$ is the bare radiative coupling of the 1s exciton \cite{kira2006many,brem2020microscopic}. It is given by 
\begin{equation}
  \gamma^{\text{exc}}= \frac{e_0^2|M_0|^2|\psi(\vec{r}=0)|^2 }{2m_{0}^2\epsilon_0cE^{\text{exc}}_0}, 
\end{equation}
where $E^{\text{exc}}$ and $\psi$ are the 1s exciton energy and wavefunctions, respectivly.

\subsubsection{Exciton-Phonon Coupling}
We take into account acoustic (longitudinal and transverse) and optical (longitudinal, transverse, and the out-of-plane A$_1$) phonons at high-symmetry points of the phononic dispersion \cite{jin2014intrinsic}, which are relevant for phonon-driven intra- and intervalley exciton scattering. For simplicity, we approximate the set of phonons at each high-symmetry point as one averaged acoustic and one averaged optical phonon \cite{lengers2020theory}. The main modes of interest for this work are the $\Gamma$ phonons, which are responsible for intravalley scattering, and $K$ phonons, which can scatter conduction electrons between the K and K' valley of the electronic bandstructure to form KK' excitons. The long-range $\Gamma$ acoustic phonon is approximated to have a linear dispersion, while all other phonons are treated with a constant energy approximation. The electron-phonon matrix elements are treated within the deformation potential approximation, with relevant parameters taken from density functional theory studies \cite{jin2014intrinsic}. Solving the Wannier equation and converting to the exciton basis then gives the exciton-phonon matrix elements, as described in previous studies \cite{selig2018dark,brem2020phonon}.

\subsubsection{Exciton-Photon Coupling In A Microcavity} \label{sec:optics}
We consider the lowest energy cavity mode of a symmetric $\lambda/2$ Fabry-Perot cavity in the Hopfield calculation. This single-mode approximation is valid due to the large free-spectral range of small microcavities. The coupling at normal incidence (i.e., $\vec{Q}=0$) of a thin excitonic material placed within the center of a high-quality and symmetric Fabry-Perot cavity can be written, close to the cavity resonance, as \cite{panzarini1999exciton,Fitzgerald2022}
\begin{equation}
    g = \hbar \sqrt{\frac{1-|r|}{|r|}\frac{\gamma^{\text{exc}}}{\tau}}, \label{eq:coupling}
\end{equation}
where $r$ is the end mirror reflection coefficient and $\tau=L/c$ is the photon travel time in a cavity of length $L$. We find a Rabi splitting of $\sim 65$ meV for a $\lambda/2$ Fabry-Perot cavity (see equation~\ref{eq:coupling}), which is in reasonable agreement with experimental works on similar systems \cite{dufferwiel2015exciton}.

Cavity TE-TM polarisation splitting is small \cite{kavokin2003thin,vasilevskiy2015exciton} and will only have a minor impact on the polariton relaxation. It is therefore ignored, with only the TM-polarized mode of the cavity considered. Idealized end mirrors are considered where both the frequency- and angle-dependence of the reflectance is disregarded, in which case, the TM coupling is constant over all momenta within the lightcone \cite{panzarini1999exciton}. In particular, this means that we neglect large-angle leakage of the Bragg mirror \cite{tassone1997bottleneck}, leading to an underestimation of the polariton radiative decay at high momenta within the lightcone.

The polariton energies and eigenfunctions (i.e, Hopfield coefficients) returned from the Hopfield diagonalization can be written down analytically for a one-exciton, one-photon system. For the lower polariton branch, they read \cite{haug2009quantum},
\begin{align}
    E_{\vec{Q}}^{\text{P}} 
    &= \frac{E^{\text{cav}}_{\vec{Q}}+E^{\text{exc}}_{\vec{Q}}}{2}-\frac{1}{2}\sqrt{(E^{\text{cav}}_{\vec{Q}}-E^{\text{exc}}_{\vec{Q}})^2+4g_{\vec{Q}}^2}
    \\
    U_{\vec{Q}} &= \sqrt{ \frac{E_{\vec{Q}}^{\text{P}} -E^{\text{exc}}_{\vec{Q}}} {2E_{\vec{Q}}^{\text{P}} -E^{\text{exc}}_{\vec{Q}}  -E^{\text{cav}}_{\vec{Q}} }}
    \\
    V_{\vec{Q}} &= \sqrt{ \frac{E_{\vec{Q}}^{\text{P}} -E^{\text{cav}}_{\vec{Q}}} {2E_{\vec{Q}}^{\text{P}} -E^{\text{exc}}_{\vec{Q}}  -E^{\text{cav}}_{\vec{Q}} }}
    ,
\end{align}
where $E^{\text{cav}}$ is the cavity energy, and the Hopfield coefficients, $U/V$, give the photonic/excitonic character of the exciton polariton.

A peculiarity of the model is that there is a small deviation between the exciton and polariton dispersion outside the lightcone, which can be seen in figure 1(b) of the main text. This is a consequence of coupling between excitons and guided waves of the TMD \cite{vasilevskiy2015exciton}. For $Q>\omega/c$, the associated wavevector perpendicular to the plane of the TMD becomes imaginary and, rather than a standing wave, the electric field exponentially decays either side of the monolayer. While in this case the impact of these non-radiative cavity modes is negligible, this does illustrate that in principle both radiative and guided modes may be important for polariton relaxation.


\subsubsection{Polariton Dynamics}

Modeling exciton-polaritons poses a numerical challenge due to the large range of momentum scales that need to be addressed, spanning from photonic phenomena on the scale of micrometers to excitonic phenomena at the nanometer scale. We have found that discretizing the Boltzmann equation using a logarithmic momentum grid works well at accurately describing both polaritons, and exciton dynamics outside the lightcone. Because we utilize a parabolic band approximation for momenta around high-symmetry points of the excitonic bandstructure, it is appropriate to take an isotropic polariton distribution, $N_{n\vec{Q}}(t) \approx N_{n Q}(t)$. The numerical evaluation can then proceed as detailed elsewhere for the excitonic Boltzmann equation \cite{selig2018dark,selig2018exciton}. Particle conservation is not enforced and can be used as a check of the numerical calculation: $2\pi\sum_n \int dQ \ Q  N_{nQ}=1$. Note that while polariton population is lost due to radiative recombination within the lightcone, the number of bright states is extremely small compared to the total KK exciton reservoir. This means that the total polariton number is approximately conserved over the time span we explore (e.g., see figure~\ref{fig:supp_2}(a)). Phonon-induced scattering to and from the upper polariton branch is included in our calculations, but is found to have a negligible contribution to the dynamics. To mimic non-resonant excitation of the system \cite{kavokin2003thin}, all dynamical calculations are initialized with a Gaussian distribution centered at $50$ meV in the KK exciton reservoir of the lower polariton branch. We have confirmed that the initialization energy does not qualitatively impact the results discussed in the main text.

\subsection{Polaritonic Boltzmann Equation} 
Our theoretical approach treats the interaction between excitons and cavity photons non-perturbatively, i.e., we perform a Hopfield diagonalization and work in the polariton basis. In contrast, interactions with phonons and the external photon reservoir are treated as weak interactions, in other words, polaritons are approximate eigenmodes of the system and scatter only weakly with the environment. The relevant microscopic Hamiltonian is given by \cite{Fitzgerald2022,ferreira2022signatures}
\begin{align}
    \hat{H} &= \sum_{n \vec{Q}}^{N}E_{n\vec{Q}}^{\text{P}}\hat{P}_{n\vec{Q}}^{\dagger}\hat{P}_{n\vec{Q}}
    +\sum_{\alpha \vec{q}}\ E_{\alpha \vec{q}}^{\text{phn}}\hat{b}_{\alpha \vec{q}}^{\dagger}\hat{b}_{\alpha \vec{q}}
    +\sum_{\vec{Q},i,\omega}\ \hbar\omega_{\vec{Q}} \hat{\mathcal{R}}_{\vec{Q},i,\omega}^{\dagger}\hat{\mathcal{R}}_{\vec{Q},i,\omega}\nonumber\\
    &+
   \sum_{n\vec{Q},m\vec{Q}',\alpha \vec{q}} \tilde{D}_{\alpha,n\vec{Q},m\vec{Q}'} \hat{P}_{n,\vec{Q}'}^{\dagger}\hat{P}_{m\vec{Q}}\left(\hat{b}_{\alpha \vec{q}}+\hat{b}_{\alpha,-\vec{q}}^{\dagger}\right)
    +
    \sum_{n\vec{Q},i,\omega}\ G_{n \vec{Q},i} \left[\hat{\mathcal{R}}_{\vec{Q},i,\omega}^{\dagger}\hat{P}_{n\vec{Q}}+\hat{\mathcal{R}}_{\vec{Q},i,\omega}\hat{P}_{n\vec{Q}}^{\dagger}\right],
\end{align}
where $\vec{q}=\vec{Q}-\vec{Q}'$ is the momentum imparted by the phonon, $\tilde{D}$ is the polariton-phonon scattering matrix element, and $\hat{P}$, $\hat{b}$ and $\hat{\mathcal{R}}$ are field operators describing polariton, phonon, and extracavity photons, respectively. The latter are a consequence of separately quantizing the internal cavity mode and the external radiation fields. This quasimode approximation is appropriate for high-quality microcavities. Polaritons can couple to the photonic reservoir via the non-zero transmission of the cavity mirrors. This is described with a real coupling term, $ G_{n \vec{Q}}$, which is taken to have a flat frequency dependence within the first Markov approximation \cite{gardiner1985input}. The index $i$ labels the port, which for a Fabry-Perot cavity corresponds to the left- and right-hand mirrors. In the case of a symmetric cavity, the emission into the two ports must be equal. This means that we can neglect the port index and multiply the resulting radiative decay rate by two.

It is convenient to distinguish the coherent and incoherent contribution to the polariton occupation
\begin{equation}
\braket{\hat{P}_{n\vec{Q}}^\dagger \hat{P}_{n\vec{Q}}} = \underbrace{\braket{\hat{P}_{n\vec{Q}}^{\dagger}}\braket{\hat{P}_{n\vec{Q}}}}_{\text{Coherent}}
+
\underbrace{\braket{\hat{P}_{n\vec{Q}}^\dagger \hat{P}_{n\vec{Q}}}_{\text{corr}}}_{\text{Incoherent}}. \label{eq:Hamiltonian}
\end{equation}
As we consider non-resonant excitation, we assume that all coherence of the system has been lost: $\braket{\hat{P}}=\braket{\hat{b}}=\braket{\hat{\mathcal{R}}}=0$. This means that the polariton occupation is given by the purely correlated part, $N_{n\vec{Q}}=\braket{\hat{P}_{n\vec{Q}}^\dagger \hat{P}_{n\vec{Q}}}=\braket{\hat{P}_{n\vec{Q}}^\dagger \hat{P}_{n\vec{Q}}}_{\text{corr}}$. Following the same prescription as used for bare excitons \cite{brem2020phonon,selig2018exciton}, the Heisenberg equation of motion is used to derive the temporal evolution of the polariton occupation yielding
\begin{equation}
    \partial_{t}N_{n\vec{Q}}(t)
    = - \frac{2 \times 2}{\hbar}\sum_{\omega}\Im\left[G_{n\vec{Q}}\mathcal{S}_{n\vec{Q},\omega}(t)\right]
    +
    \frac{2}{\hbar}\sum_{m\vec{Q}',\alpha,\pm} \Im\left[
    \tilde{D}_{\alpha,n\vec{Q},m\vec{Q}'}\mathcal{C}_{\alpha,n\vec{Q},m\vec{Q}'}^{\pm}(t)\right] \label{eq:equation_of_motion}
\end{equation}
where $\mathcal{C}_{\alpha,n\vec{Q},m\vec{Q}'}^{\pm}=\braket{\hat{P}_{n\vec{Q}}^{\dagger}\hat{P}_{m\vec{Q}'}\hat{b}_{\alpha,\mp\vec{q}}^{(\dagger)}}_{\text{corr}}$ describes polariton-phonon correlation for phonon emission/absorption ($\pm$), and $\mathcal{S}_{n\vec{Q},\omega}=\braket{\hat{\mathcal{R}}_{\vec{Q}\omega}^{\dagger}\hat{P}_{n\vec{Q}}}_{\text{corr}}$ is the polariton radiative-recombination amplitude.

\subsubsection{Polariton-Phonon Dynamics}
The dynamics driven by the polariton-phonon interaction is found by truncating the many-particle hierarchy on the level of two-particle correlations (e.g., ignoring two-phonon processes) and assuming low polariton densities \cite{brem2020phonon,kira2006many}
\begin{equation}
   i\hbar\partial_{t}\mathcal{C}_{\alpha,n\vec{Q},m\vec{Q}'}^{\pm}(t)|_{\text{P-Phn}}=\left(E_{m\vec{Q}'}^{\text{P}}-E_{n\vec{Q}}^{\text{P}}\pm E_{\vec{q}}^{\text{phn}}\right)\mathcal{C}_{\alpha,n\vec{Q},m\vec{Q}'}^{\pm}(t)
   +
  \tilde{D}_{\alpha,n\vec{Q},m\vec{Q}'}\left(N_{n\vec{Q}}\eta_{\vec{q}}^{\pm}-N_{m\vec{Q}'}\eta_{\vec{q}}^{\mp}\right),
\end{equation}
where the phonon occupation factor, $\eta_{\vec{q}}^{\pm}=\frac{1}{2}\pm\frac{1}{2}+n_{\vec{q}}^{\text{phn}}$, has been introduced. This first-order differential equation in time is now solved within the Markov approximation, yielding
\begin{equation}
    \mathcal{C}_{\alpha,n\vec{Q},m\vec{Q}'}^{\pm} \approx
    -i\pi
    \tilde{D}_{\alpha,n\vec{Q},m\vec{Q}'}^{*}
    \left(N_{n\vec{Q}}\eta_{\vec{q}}^{\pm}-N_{m\vec{Q}'}\eta_{\vec{q}}^{\mp}\right)\delta\left(E^{\text{P}}_{m \vec{Q}'}-E^{\text{P}}_{n\vec{Q}}\pm E^{\text{phn}}_{\mp \vec{q}}\right), \label{eq:phonon_markov}
\end{equation}
where we only take into account the dephasing effect of polariton-phonon scattering and ignore the polaron shift. The Dirac delta function strictly enforces energy and momentum conservation of the polariton-phonon scattering. This leads to the abrupt opening of scattering channels at certain momenta of the polariton dispersion. Note that the neglected higher-order dephasing processes are expected to broaden the opening of scattering channels. 

Substitution of equation \ref{eq:phonon_markov} into equation \ref{eq:equation_of_motion} gives
\begin{equation}
    \partial_tN_{n\vec{Q}}|_{\text{P-phn}} = 
    \sum_{m\vec{Q}'} \left(W_{m\vec{Q}',n\vec{Q}}N_{m\vec{Q}'}(t)-W_{n\vec{Q},m\vec{Q}'}N_{n\vec{Q}}(t)\right), 
\end{equation}
where the scattering matrix is given by summing over all possible scattering phonon emission and absorption channels
\begin{equation}
    W_{n\vec{Q},m\vec{Q}'}=\frac{2\pi}{\hbar}\sum_{\alpha,\pm}
    \left|\tilde{D}_{\alpha,n\vec{Q},m\vec{Q}'} \right|^{2}
\left(\frac{1}{2}\pm\frac{1}{2}+n_{\alpha,|\vec{Q}-\vec{Q}'|}^{\text{ph}}\right) 
 \delta\left(E_{m\vec{Q}'}-E_{n\vec{Q}} \pm E^{\text{phn}}_{\alpha,|\vec{Q}-\vec{Q}'|}\right).
\end{equation}
It describes the probability rate for a phonon to scatter an exciton from state $\ket{n,\vec{Q}}$ to $\ket{m,\vec{Q}'}$. For the out-scattering term, the occupation can be taken out the summation over the final state, and the out-scattering matrix can be linked to the phonon-induced dephasing rate, $\Gamma_{n\vec{Q}}= \sum_{m\vec{Q}'}W_{n\vec{Q},m\vec{Q}'}/2$, i.e., Eq.~(4) of the main text.

\subsubsection{Polariton Decay Dynamics}
In a similar fashion, the dynamics of the polariton radiative decay can be found using the Heisenberg equation of motion 
\begin{equation}
i\hbar\partial_{t}\mathcal{S}_{n\vec{Q},\omega}(t)=\left(E^{\text{P}}_{n\vec{Q}}-\hbar\omega_{\vec{Q}}\right)\mathcal{S}_{n\vec{Q},\omega}(t)-G_{n\vec{Q}}^{*}N_{n\vec{Q}}(t).
\end{equation}
Again, this can be solved within the Markov approximation, yielding
\begin{equation}
    \mathcal{S}_{n\vec{Q},\omega}(t)\approx i\pi G_{n\vec{Q}}^{*}N_{n\vec{Q}}(t)\delta\left(E^{\text{P}}_{n\vec{Q}}-\hbar\omega_{\vec{Q}}\right).
\end{equation}
Substitution into equation \ref{eq:equation_of_motion} then gives \cite{portolan2008nonequilibrium,Fitzgerald2022}
\begin{align}
    \partial_{t}N_{n\vec{Q}}(t)|_{\text{P-pht}} &= -\frac{4}{\hbar}\sum_{\omega}\Im\left[G_{n\vec{Q}}\mathcal{S}_{n\vec{Q},\omega}(t)\right] \nonumber
    \\
    &= 
    -\frac{4\pi}{\hbar} \sum_{\omega}|G_{n\vec{Q}}|^{2}\delta\left(\omega^{\text{P}}_{n\vec{Q}}-\omega_{\vec{Q}}\right)N_{n\vec{Q}}(t) \nonumber
    \\
    &=-2\gamma_{n\vec{Q}}N_{n\vec{Q}}(t)
\end{align}
Here, $\gamma_{n\vec{Q}}$ is the total radiative decay rate of polaritons, i.e., the sum of the photon leakage through both ports. Intuitively, it is given by the total cavity decay rate scaled by the photonic Hopfield coefficient. For a high-quality and symmetric cavity, we find \cite{Fitzgerald2022}, 
\begin{equation}
    \gamma_{n\vec{Q}} =\frac{2\pi}{\hbar}|G_{n\vec{Q}}|^2 = \frac{cT|U_{n\vec{Q}}|^2}{2L},
\end{equation}
where $L$ is the cavity length and $T=1-|r|^2$ is the end mirror transmission.

\subsection{Polariton Photoluminescence} 
In experiments, photoluminescence (PL) gives access to the polariton dispersion and the polariton occupation within the lightcone. An equation of motion approach is used for the purely correlated part of the extracavity photon density, 
\begin{equation}
N_{\vec{Q}\omega}^{\mathcal{R}}=\braket{\hat{\mathcal{R}}_{\vec{Q}\omega}^{\dagger}\hat{\mathcal{R}}_{\vec{Q}\omega'}}=\braket{\hat{\mathcal{R}}_{\vec{Q}\omega}^{\dagger}\hat{\mathcal{R}}_{\vec{Q}\omega'}}_{\text{corr}}.
\end{equation}
We solve this equation for a single port and multiply the final result by two. The coupled equations of motion for the external photonic density and polariton recombination read
\begin{align}
\partial_{t}N_{\vec{Q}\omega}^{\mathcal{R}}(t)
&=\frac{2\times 2}{\hbar}\sum_{n}\Im\left[G_{n\vec{Q}}\mathcal{S}_{n\vec{Q},\omega}(t)\right]
\label{eq:PL_dynamics}
\\
i\hbar\partial_{t}\mathcal{S}_{n\vec{Q},\omega}
&=
\left(E_{n\vec{Q}}^{\text{P}}-\hbar\omega_{\vec{Q}}\right)\mathcal{S}_{n\vec{Q},\omega}(t)
- G_{n\vec{Q}}^*N_{n\vec{Q}}(t)
+
  \sum_{\omega} G_{n\vec{Q}}^* N_{\vec{Q}\omega}^{\mathcal{R}}(t)  \nonumber
\\
&+ \sum_{m,\alpha \vec{q}, \pm} \tilde{D}_{\alpha,n\vec{Q},m\vec{Q}} \mathcal{U}_{n\vec{Q},\alpha\vec{q}}^{\pm},
\end{align}
where $\mathcal{U}_{n\vec{Q},\alpha\vec{q}}^{\pm} = \braket{\hat{\mathcal{R}}_{\vec{Q}\omega}^{\dagger} \hat{P}_{n,\vec{Q}-\vec{q}} \hat{b}_{\alpha,\mp \vec{q}}^{(\dagger)} }_{\text{corr}}$ corresponds to phonon-assisted polariton decay via phonon emission/absorption. Here, we have neglected coupling between different polariton branches \cite{Fitzgerald2022}, i.e., we approximate that the recombination dynamics of the $n$th polariton only depends on the occupation of that polariton branch: $\dot{\mathcal{S}_n}\propto \mathcal{S}_m \delta_{nm}$. Using the Markov approximation and disregarding energy renormalization terms, the third and fourth term can be shown to lead to the radiative decay and phonon-induced dephasing rates 
\begin{equation}
i\hbar\partial_{t}\mathcal{S}_{n\vec{Q},\omega} \approx \left(E_{n\vec{Q}}^{\text{P}}-\hbar\omega_{\vec{Q}}-i\hbar(\Gamma_{n\vec{Q}}+\gamma_{n\vec{Q}})\right)\mathcal{S}_{n\vec{Q},\omega}(t)
- G_{n\vec{Q}}^*N_{n\vec{Q}}(t). 
\end{equation}
The dynamics for $\mathcal{S}$ are then solved in the static limit, $\partial_{t}\mathcal{S}=0$, and then substituted into equation \ref{eq:PL_dynamics}. Finally, the PL is related to the temporal change of the photon population \cite{kira2006many,selig2018exciton}
\begin{equation}
I(\omega,t) =\partial_{t}N_{\vec{Q}\omega}^{\mathcal{R}}(t) \propto 
\frac{\gamma_{n\vec{Q}}(\gamma_{n\vec{Q}}+\Gamma_{n\vec{Q}})}{\left(\omega_{n\vec{Q}}^{\text{P}}-\omega_{Q}\right)^{2}+(\gamma_{n\vec{Q}}+\Gamma_{n\vec{Q}})^{2}}N_{n\vec{Q}}(t),
\end{equation}
which is the \emph{polaritonic PL Elliot formula} for a single branch (assuming energetically well-separated branches relative to the polariton linewidth). It has an identical structure to the regular excitonic PL Elliot formula \cite{selig2018exciton,brem2020phonon}. When evaluated at the polariton resonance, $\hbar\omega_{Q} = E_{n\vec{Q}}^{\text{P}}$, equation (3) of the main text is found.

\begin{figure}[b!]
\includegraphics[width=\columnwidth*3/5]{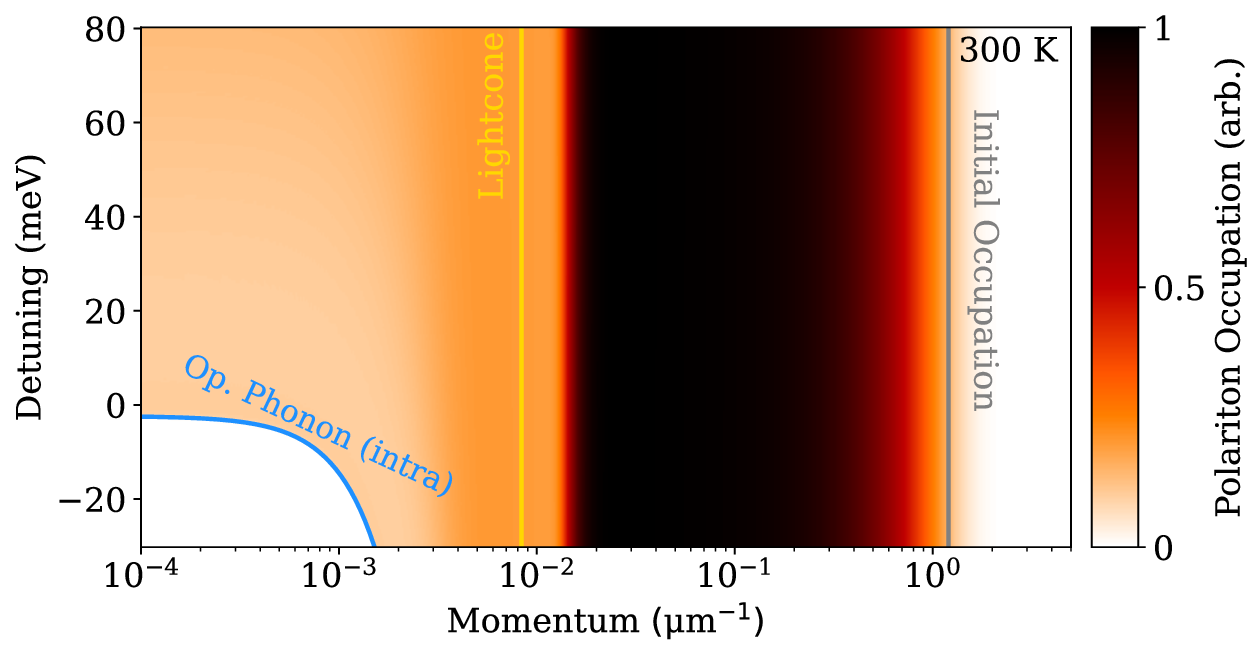}
\caption{Detuning dependence of the momentum-resolved steady-state lower polariton occupation at $300$ K where intervalley scattering is forbidden.
\label{fig:supp_1}}
\end{figure}

\subsection{Steady-State Occupation Depletion In the Presence of Radiative Recombination}
The separation of the LP dispersion into an excitonic and polariton domain, combined with weak phonon-driven scattering within the lightcone, means that we can derive a simple expression for the steady-state solution of the polaritonic Boltzmann equation in the presence of radiative decay. At steady-state, equation (1) of the main text can be written as
\begin{equation}
    \sum_{m,\vec{Q}'}W_{m\vec{Q}',n\vec{Q}} N_{m\vec{Q}'}(t_\infty) = 2(\Gamma_{n\vec{Q}} + \gamma_{n\vec{Q}})N_{n\vec{Q}}(t_\infty). \label{eq:steady_state_2}
\end{equation}
In the absence of radiative decay, the steady-state occupation is given by a Boltzmann distribution, $N^{0}_{\vec{Q}}$. This means
\begin{equation}
    \sum_{m,\vec{Q}'}W_{m\vec{Q}',n\vec{Q}} N^{0}_{m\vec{Q}'} = 2\Gamma_{n\vec{Q}} N^{0}_{n\vec{Q}}. \label{eq:steady_state_3}
\end{equation}
We now make two assumptions: 1) outside the lightcone the steady-state solution is thermalized, and 2) the vast majority of states that contribute to the in-scattering are outside the lightcone. This means we can state
\begin{equation}
 \sum_{m,\vec{Q}'}W_{m\vec{Q}',n\vec{Q}} N_{m\vec{Q}'}(t_\infty) \approx  \sum_{m,\vec{Q}'}W_{m\vec{Q}',n\vec{Q}} N^{0}_{m\vec{Q}'},   
\end{equation}
which immediately leads to equation (2) of the main text when combining equations \ref{eq:steady_state_2} and \ref{eq:steady_state_3}. 

It is interesting to note that equation (2) of the main text can be rewritten in the following illustrative form
\begin{equation}
    N_{n\vec{Q}}(t_\infty) = \frac{\tilde{\Gamma}^{\text{in}}_{n\vec{Q}}}{\Gamma_{n\vec{Q}} + \gamma_{n\vec{Q}}},
\end{equation}
which is the ratio of the approximate total in-scattering, $\tilde{\Gamma}^{\text{in}}_{n\vec{Q}}=\sum_{m,\vec{Q}'}W_{m\vec{Q}',n\vec{Q}} N^{0}_{m\vec{Q}'}/2$ and total out-scattering rate for the state $\ket{n,\vec{Q}}$.

\subsection{Steady State Occupation in the Absence of Intervalley Scattering}
To further illustrate the crucial impact that intervalley scattering can play in the thermalization of exciton polaritons, in figure~\ref{fig:supp_1} the detuning dependence of the steady-state LP occupation is shown where all intervalley scattering channels are artificially switched off, and for a temperature of $300$ K.  In contrast to figure~2(b) of the main text, once the intravalley optical phonon channel is open (blue curve) there is very little detuning dependence of the occupation within the lightcone (gold vertical line). In the absence of any intervalley scattering channels, at all detunings the LP occupation within the entire lightcone is depleted relative to the occupation outside, and is much smaller than the corresponding thermalized Boltzmann distribution. The polariton bottleneck is located outside the lightcone and indicated by the abrupt change of occupation. This is a consequence of the assumption of perfect mirrors with angle-independent reflectance (see section \ref{sec:optics}).

\subsection{Dynamics of the Dark KK' Exciton Reservoir}

\begin{figure}[h!]
\includegraphics[width=\columnwidth*4/5]{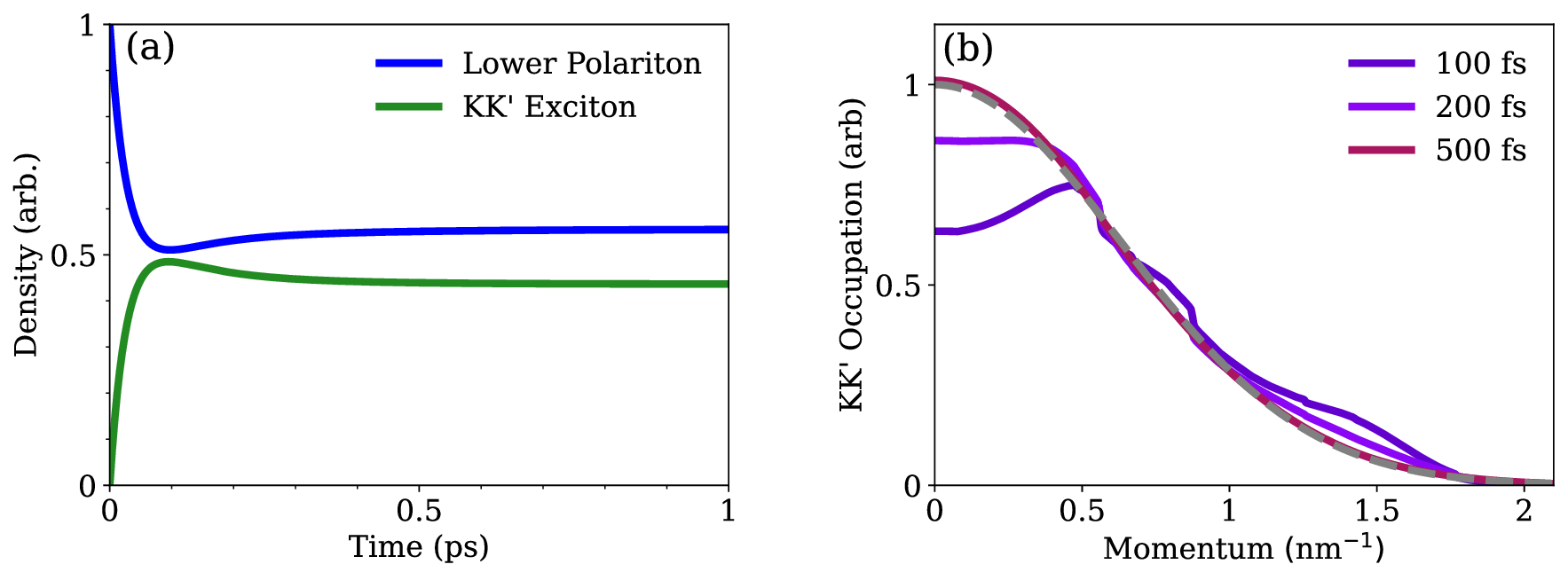}
\caption{
 (a)~Time dependence of the lower polariton and the dark KK' exciton density at $300$ K and for a cavity detuning of $\Delta=20$ meV, which corresponds to the opening of the optical intervalley scattering channel for the polariton ground state. (b)~Snapshots of the momentum-resolved KK' exciton occupation at fixed times. The dashed grey line gives the Boltzmann distribution.
\label{fig:supp_2}}
\end{figure}

To complement the discussion of figure~3 in the main text, we show in figure~\ref{fig:supp_2}(a) the time-resolved dynamics of the LP and dark KK' density, $\sum_{\vec{Q}}N_{\vec{Q}}(t)$, at a temperature of $300$ K and a cavity detuned to the optical phonon scattering channel  (i.e., $\Delta=20$ meV, see figure~1(b) in the main text). There is a rapid transfer of density from the initially populated KK exciton reservoir of the LP branch to KK' exciton states over tens of fs, in agreement with previous studies of bare excitons in \ce{MoSe2} \cite{selig2018dark}. The density reaches an equilibrium in less than $500$ fs, indicating a rapid thermalization of both the KK and KK' exciton reservoirs via the efficient phonon-assisted scattering at elevated temperatures. This result shows that, because of its close energetic position relative to the bright KK exciton, the KK' exciton valley has a significant occupation at room temperature and thus can act as an efficient reservoir for populating polariton states.

These observations are further confirmed in figure~\ref{fig:supp_2}(b), where the time evolution of the momentum-resolved KK' exciton occupation is shown. In under $100$ fs there is a rapid intervalley transfer of population from the initial distribution in the KK reservoir to KK' excitons via K phonon emission. There is then an efficient redistribution of KK' occupation towards a thermalized Boltzmann distribution (grey dashed line) via intravalley scattering on a timescale of about $500$ fs.

\subsection{Cavity Tuning Of Polariton-Enhanced Phonon Sidebands}

\begin{figure}[t!]
\includegraphics[width=\columnwidth*3/5]{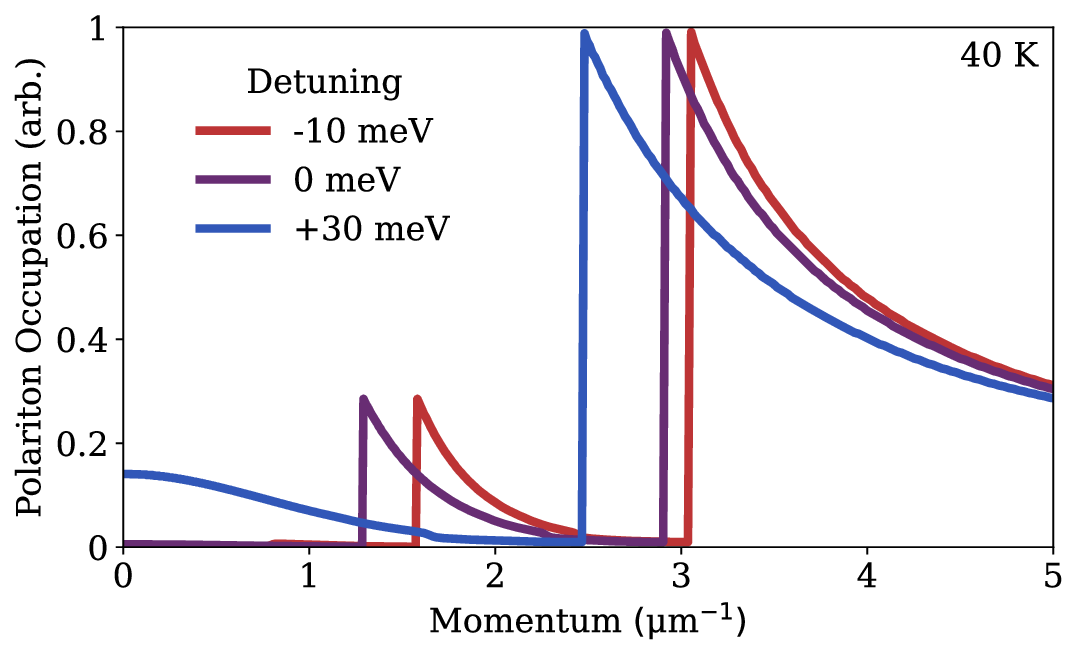}
\caption{Detuning dependence of the lower polariton occupation at $10$ ps and a temperature of $40$ K.
\label{fig:supp_3}}
\end{figure}

 In figure~\ref{fig:supp_3}, the momentum-dependent LP occupation is shown within the lightcone, evaluated at the fixed time of $10$ ps and a temperature of $40$K for the same three representative detunings as used in figure~2(a) of the main text. Similarly to the room temperature case, we observe detuning-dependent, sharp changes in occupation corresponding to the opening of the intervalley scattering channels. In contrast, the LP occupation now does not follow a thermalized Boltzmann distribution when the channels open, but instead possesses a distinctive asymmetric lineshape with a high momentum tail. For the red-detuned case ($\Delta=-10$ meV, red line), the opening of both intervalley channels is visible, with the peak corresponding to the optical channel at $Q=1.6 \ \mu\text{m}^{-1}$ and the larger acoustic peak at $Q=3 \ \mu\text{m}^{-1}$. The opening of the intravalley channel is barely visible at this low temperature because of the small exciton-phonon matrix element. As the cavity is blue detuned, the peaks shift to lower momenta within the lightcone (purple and blue lines). For the blue-detuned cavity ($\Delta=30$ meV, blue line), the intra- and intervalley optical channel is open at all momenta in the LP dispersion, leading to a small occupation at the ground state. This is significantly reduced compared to the situation at $300$ K (figure~2(a) of the main text). The weaker in-scattering is due to the diminished occupation of the KK' reservoir and a reduced exciton-phonon scattering at low temperatures.

\end{document}